\documentclass[aps,prd,10pt,showpacs,showkeys,amsmath,amssymb,twocolumn,superscriptaddress,groupedaddress]{revtex4-1}
\usepackage{epsfig}
\usepackage{graphicx}
\usepackage{graphics}
\usepackage{xspace}
\usepackage{amssymb}
\usepackage{amsmath}
\usepackage[dvips]{color}
\usepackage{latexsym}
\usepackage{mathrsfs}
\usepackage{url}
\usepackage[english]{babel}
\newcommand{\inieq}{\begin{eqnarray}}            
\newcommand{\fineq}{\end{eqnarray}}            
\newcommand{\be}{\begin{equation}}
\newcommand{\ee}{\end{equation}}
\newcommand{\ba}{\begin{eqnarray}}
\newcommand{\ea}{\end{eqnarray}}

\def\ee{\mbox{$\left(e,e^{\prime}\right)$\ }}
\def\eep{\mbox{$\left(e,e^{\prime}p\right)$\ }}

\begin{document}
\title{Relativistic descriptions of final-state interactions
in charged-current quasielastic antineutrino-nucleus scattering at MiniBooNE
kinematics.}
\author{Andrea Meucci} 
\affiliation{Dipartimento di Fisica Nucleare e Teorica, 
Universit\`{a} degli Studi di Pavia and \\
INFN,
Sezione di Pavia, via A. Bassi 6, I-27100 Pavia, Italy}
\author{Carlotta Giusti}
\affiliation{Dipartimento di Fisica Nucleare e Teorica, 
Universit\`{a} degli Studi di Pavia and \\
INFN,
Sezione di Pavia, via A. Bassi 6, I-27100 Pavia, Italy}

\date{\today}

\begin{abstract}
The analysis of the recent charged-current neutrino-nucleus
scattering cross sections measured by the  MiniBooNE Collaboration requires relativistic 
theoretical descriptions also accounting for the role of final-state 
interactions. In this work, we evaluate differential antineutrino-nucleus
cross sections with  
the relativistic Green's function model, where the final-state interactions are described in 
the inclusive scattering consistently with the exclusive scattering using a complex 
optical potential.
The sensitivity to the parameterization adopted for the 
phenomenological optical potential is discussed. The predictions of the 
relativistic Green's function model are compared with the results of
different descriptions of final-state interactions.
 \end{abstract}


\maketitle


\section{Introduction}
\label{intro}

The MiniBooNE Collaboration has recently reported~\cite{miniboone} a 
measurement of the charged-current quasielastic (CCQE) flux-averaged double-differential 
muon neutrino cross section
on $^{12}$C in an energy range up to  
$\approx 3\ $GeV. The neutrino-nucleus CCQE reaction in MiniBooNE 
may be considered as scattering of an incident 
neutrino with a single nucleon bound in carbon, but 
it can also be sensitive to contributions from collective nuclear 
effects, whose clear understanding is crucial for the analysis of ongoing 
and future  neutrino oscillation 
measurements~\cite{miniboone,miniboonenc,Nakajima:2010fp,Gran:2006jn,Wu08,
Lyubushkin:2008pe,minos10}.  

When a dipole dependence on the four-momentum
transferred squared $Q^2$ is assumed for the axial form factor, the nucleon axial mass 
$M_A$  has been used as a free parameter within the relativistic Fermi gas 
(RFG) model~\cite{Casper:2002sd,Hayato:2002sd}.
Recent CCQE
measurements~\cite{Gran:2006jn,Nakajima:2010fp} reported values of 
$M_A \approx \ 1.2$ GeV/$c^2$, significantly larger than the 
world average value from the deuterium data of 
$M_A = 1.03$ GeV/$c^2$~\cite{Bernard:2001rs,bodek08}. 
In agreement with these results, the MiniBooNE collaboration reported values of 
$M_A =\ 1.35 \pm 0.17$ GeV/$c^2$ for the CCQE measurements~\cite{miniboone} and 
$M_A = \ 1.39 \pm 0.11$ GeV/$c^2$ for the neutral-current elastic (NCE) 
data~\cite{miniboonenc}. 
A recent application of analyticity and dispersion relations to the axial vector form 
factor to find constraints for the axial mass parameter using the CCQE data from 
MiniBooNE is presented in Ref.~\cite{bhatta} and produces a value of
$M_A=0.85^{+0.22}_{-0.07} \pm {0.09} $  GeV, which significantly differs from the RFG
extraction.

The energy region 
considered in the MiniBooNE experiments, with neutrino and antineutrino energy up to 
$\approx \ 3$ GeV and average energy of the muon neutrino (antineutrino) flux 
$\approx \ 0.79\ (0.66)$ GeV~\cite{AguilarArevalo:2008yp}, requires the use of a relativistic model, where 
not only relativistic kinematics should be considered, but also nuclear 
dynamics and current operators should be described within a relativistic 
framework. From the comparison with electron scattering data it is known that the 
RFG, although able of getting the basic shape and size of the response, 
turns out to be a too naive model to correctly account for important details of the nuclear 
dynamics. Thus, the larger axial mass needed by the RFG could be considered as 
an effective value to incorporate nuclear effects into the calculation rather than a 
clear signal of a modified axial mass.

At intermediate energy, quasielastic (QE) electron scattering 
calculations~\cite{Boffi:1993gs,book}, which were able to
successfully describe a wide number of experimental data, can provide a 
useful tool to study neutrino-induced processes. 
Several theoretical models have been applied in recent years to
$\nu$-nucleus scattering reactions and some of them have been compared with 
the MiniBooNE data, both in the CCQE and in the NCE channels. 
At the level of the impulse approximation (IA), 
models based on the use of a realistic spectral 
function~\cite{Benhar:2010nx,Benhar:2011wy}, 
which are built within a nonrelativistic framework, underestimate the
experimental CCQE and NCE cross sections unless $M_A$ is enlarged with respect
to the world average value. 
The same results are obtained by models based on the relativistic IA 
(RIA)~\cite{Butkevich:2010cr,Butkevich:2011fu,
jusz10}.  
However, the reaction may have significant contributions from effects beyond 
the IA in some kinematic regions where the experimental neutrino flux has significant 
strength.  
For instance, in the models of Refs.~\cite{Martini:2009uj,Martini:2010ex,Martini:2011wp,
Nieves:2011pp,Nieves:2011yp}
the contribution of multinucleon excitations to CCQE scattering has been 
found sizable and able to bring the theory in agreement with the experimental
MiniBooNE cross sections without increasing the value of $M_A$. 

The role of processes involving two-body currents compared to the IA models
has been discussed in Refs.~\cite{Benhar:2011wy,Amaro:2010sd,Amaro:2011qb,bodek11}. 
A careful evaluation of all nuclear effects 
and of the relevance of multinucleon emission and of some non-nucleonic 
contributions~\cite{PhysRevC.79.034601,Leitner:2010kp,
PhysRevC.83.054616,FernandezMartinez2011477} 
would be interesting for a deeper
understanding of the reaction dynamic.  
However, fully relativistic microscopic calculations of two-particle-two-hole (2p-2h) 
contributions are extremely difficult and may be bound to model dependent assumptions. 
For instance, the part of the 2p-2h 
excitations which may be reached through two-body meson-exchange 
currents (MEC), in particular the contribution of the vector MEC in the 
2p-2h sector, evaluated in the model of Ref.~\cite{DePace:2003xu}, has been 
incorporated in a phenomenological approach based on the superscaling behavior 
of electron scattering data~\cite{Amaro:2010sd,Amaro:2011qb,Amaro:2011aa}. 
The effects of MEC are important relative to the QE contribution, especially for the 
antineutrino cross section, where the destructive vector-axial interference term 
reduces the pure QE contribution and the MEC have a more significant 
role~\cite{Amaro:2011aa}.

Within the QE kinematic domain, the treatment of the final-state 
interactions (FSI) between the ejected nucleon and the residual nucleus is an 
essential ingredient for the comparison with data. The relevance of FSI has been
clearly stated in the case of exclusive \eep processes, where the 
use of  complex optical potentials in the distorted wave impulse approximation 
(DWIA) is required~\cite{Boffi:1993gs,book,Udias:1993xy,Udias:1993zs,Meucci:2001qc,
Meucci:2001ty,Radici:2003zz,Tamae:2009zz,Giusti:2011it}. 
However, the pure DWIA approach, which is based on the use of an 
absorptive complex potential, would be inconsistent in the 
analysis of 
inclusive scattering, where all final-state channels should be retained and the total
flux, although redistributed among all possible channels due to FSI, must be conserved.
Different approaches have been used to describe FSI in relativistic calculations for 
the inclusive QE electron- and neutrino-nucleus 
scattering~\cite{Maieron:2003df,Caballero:2006wi,Caballero:2009sn,Meucci:2003cv,
Meucci:2003uy,Meucci:2004ip,Meucci:2006cx,Meucci:2006ir,Meucci:2008zz,
Meucci:2009nm,Meucci:2011pi}. 
In the relativistic plane-wave impulse approximation (RPWIA), FSI 
are simply neglected. In another approach, FSI are included in DWIA calculations 
where the 
final nucleon state is evaluated with real potentials, either retaining 
only the real part of the relativistic energy-dependent complex optical potential 
(rROP), or using the same relativistic mean field potential considered in 
describing the initial nucleon state (RMF). 
Although conserving the flux, the rROP is unsatisfactory from a theoretical 
point of view, since it relies on  an energy-dependent potential, which 
reflects the different contribution of open inelastic channels for each energy, 
and under such conditions dispersion relations dictate that the potential 
should have a nonzero imaginary term~\cite{hori}.
On the other hand, in the RMF model the same strong energy-independent real 
potential is used for both bound and scattering states. It fulfills the 
dispersion relations~\cite{hori} and also the continuity equation. 

In a different description of FSI relativistic Green's function (RGF) 
techniques~\cite{Capuzzi:1991qd,Meucci:2003uy,Meucci:2003cv,Capuzzi:2004au,
Meucci:2009nm,Meucci:2011pi}
are used. 
In the RGF model, under suitable approximations, which are basically related 
to the impulse approximation, the components of the hadron tensor are written 
in terms of the single particle optical model Green's function, whose 
self-energy is the Feshbach optical 
potential. The explicit calculation of the single particle Green's function 
can be avoided by its spectral representation, which is based on a biorthogonal 
expansion in terms of a non Hermitian optical potential $\cal H$ and of its 
Hermitian conjugate $\cal H^{\dagger}$. Calculations require matrix elements 
of the same type as the DWIA ones for the case of exclusive \eep processes in 
Ref.~\cite{Meucci:2001qc}, but involve 
eigenfunctions of both $\cal H$ and $\cal H^{\dagger}$, where the imaginary 
part gives in one case an absorption and in the other case a gain of flux.
This formalism allows us to reconstruct the flux lost into nonelastic 
channels in the case of the inclusive response starting from the complex optical 
potential which describes elastic nucleon-nucleus scattering data. 
Thus, it provides a consistent treatment of FSI 
in the exclusive and in the inclusive scattering and gives a 
good description of \ee data~\cite{Meucci:2003uy,Meucci:2009nm}. 
Due to the analyticity properties of the optical potential, 
the RGF model fulfills the Coulomb sum rule~\cite{hori,Capuzzi:1991qd,Meucci:2003uy}.
In addition, the RMF and RGF reproduce also
the behavior of the scaling function extracted from the electron scattering 
data~\cite{Meucci:2009nm}.

These different descriptions of FSI have been compared in~\cite{Meucci:2009nm} 
for the inclusive QE electron scattering, in Ref.~\cite{Meucci:2011pi} for 
the CCQE neutrino scattering, and in Refs.~\cite{Meucci:2011vd,Meucci:2011nc} 
with the CCQE and NCE MiniBooNE data.
In Ref.~\cite{Meucci:2011vd} both the 
RMF and the RGF give a good description of the shape of the CCQE experimental 
data and, moreover, the RGF can give cross sections of the same magnitude as 
the experimental ones without the need to 
increase the value of $M_A$. 
Similar results are obtained in Ref.~\cite{Meucci:2011nc}, where the RGF 
results and their interpretation in comparison with the NCE data from MiniBooNE 
are discussed. 

In this paper different relativistic descriptions of FSI for CCQE $\bar \nu$-nucleus 
reactions are discussed and results for the double-differential cross section 
averaged over the $\bar \nu_{\mu}$
MiniBooNE flux are presented. The MiniBooNE collaboration has accumulated 
an extensive data set of 
$\nu_{\mu}$ events, but it has also measured $\bar \nu_{\mu}$ CCQE events.  
The analysis of the antineutrino data is currently 
ongoing~\cite{minibooneflux} and some preliminary 
results can be found in the MiniBooNE website~\cite{minibooneweb}. 
When available, the antineutrino measurements will be an additional source 
of information about the weak charged-current lepton-nucleus interaction 
and,  combined with the corresponding neutrino data, will provide  important 
insight about the role of the 
longitudinal, transverse, and interference vector-axial responses which enter the
cross sections. Indeed, being aware of the interpretative questions which may be connected to
the fact that the neutrino and the antineutrino fluxes at MiniBooNE
are different, with the $\bar\nu_{\mu}$ flux significantly smaller and with lower average energy than 
the $\nu_{\mu}$ one, measurements of both reactions would  
be useful to clarify the role of nuclear effects in the analysis 
of lepton-nucleus processes.



\section{Results and discussion}
\label{results}

In all the calculations presented in this work the bound nucleon states 
are taken as self-consistent Dirac-Hartree solutions derived 
within a relativistic mean field approach using a Lagrangian containing 
$\sigma$, $\omega$, and $\rho$ mesons~\cite{Serot:1984ey,Sharma:1993it,
Lalazissis:1996rd}. In the RGF calculations we have used two parameterizations for the 
relativistic optical potential: 
the Energy-Dependent but A-Independent EDAI (where the $A$ represents the atomic number)
and the Energy-Dependent and A-Dependent EDAD1 complex   phenomenological potentials 
of Refs.~\cite{chc,Cooper:1993nx,Cooper:2009}, which are 
fitted to elastic proton scattering data in an energy range 
up to 1040 MeV.
We note that whereas EDAD1 is a global parameterization, EDAI is a 
single-nucleus parameterization, which is 
constructed to 
fit scattering data just on $^{12}$C and, as such, does 
have an edge in terms of better reproduction of the elastic proton-$^{12}$C 
phenomenology \cite{chc} compared to EDAD1, and also leads to a better
description of the inclusive quasielastic $(e,e^{\prime})$ cross sections, as
well as to CCQE and NCE results in better agreement with  the MiniBooNE 
data~\cite{Meucci:2011vd,Meucci:2011nc}.


\begin{figure}[tb]
\begin{center}
\includegraphics[scale=0.45]{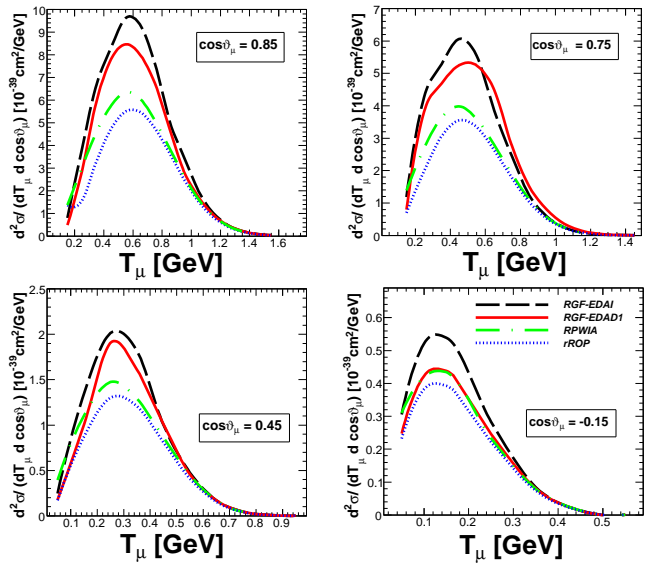} 
\end{center}
	\caption{(color online) Flux-averaged double-differential cross section per target nucleon for 
the CCQE $^{12}$C$\left(\bar\nu_{\mu}, \mu^+\right)$ reaction as a function of
$T_{\mu}$ 
for four angular bins of $\cos\vartheta_{\mu}$ calculated with the RGF-EDAD1 (solid lines) 
and the RGF-EDAI (dashed lines). The dotted lines are the rROP results calculated with 
the EDAI potential and the dot-dashed lines are the RPWIA results.	
\label{f1}}
	\end{figure}

In Fig.~\ref{f1} the CCQE double-differential 
$^{12}$C$\left(\bar\nu_{\mu}, \mu^+\right)$  cross section per target 
nucleon integrated over the MiniBooNE $\bar\nu_{\mu}$ flux 
is shown as a function 
of the muon kinetic energy $T_{\mu}$ for four angular bins of 
$\cos\vartheta_{\mu}$, where $\vartheta_{\mu}$ is the muon scattering angle, 
ranging from forward to backward angles. 
In the RPWIA calculations FSI are completely neglected. 
The rROP results, where calculations are performed with a pure real optical potential, 
are usually 15\% lower than the RPWIA ones. The rROP generally underestimates the $\nu$ 
experimental data unless a larger axial mass, e.g., $M_A \approx 1.3 - 1.4$ GeV$/c^2$,   
is used. However, 
independently of its comparison with the data, the rROP model, which is based on 
an energy-dependent potential, has important physical 
drawbacks ~\cite{hori,Meucci:2003uy,Meucci:2011vd,Meucci:2011pi}.
The RGF cross sections with both optical potentials are larger than the 
RPWIA and the rROP ones. The differences between the RGF results with the two 
optical potentials are clearly visible. For instance, the EDAI and EDAD1 potentials 
yield differences  by about $15\% - 20\%$ to the cross section in the peak region for 
the  forward angle scattering bins $\cos\vartheta_{\mu}= 0.85,\ 0.75$. Somewhat closer 
predictions are obtained in the bin $\cos\vartheta_{\mu}= 0.45$, while for the backward 
angular bin $\cos\vartheta_{\mu}= -0.15$ the differences  are enhanced up to $25\%$, but the magnitude 
of the cross sections is significantly 
reduced.  We note that the relative differences between the RGF results
with the two optical potentials are somewhat  
larger in neutrino scattering~\cite{Meucci:2011vd}.
The different behavior of the RGF in 
neutrino and antineutrino scattering is related to the relative strength of the 
vector-axial response,  which is constructive in $\nu$ scattering and
destructive in $\bar\nu$ scattering with respect to the longitudinal and 
transverse ones~\cite{Meucci:2011pi}.
Moreover, the differences between the neutrino and antineutrino MiniBooNE
fluxes, make the comparison between the results of $\nu_{\mu}$-nucleus and 
$\bar\nu_{\mu}$-nucleus scattering  not straightforward. 

The comparison between the RGF results obtained with the EDAI and EDAD1 
potentials can give an idea of how  the predictions of the model are affected 
by uncertainties in the determination of the phenomenological optical
potential. 
The differences depend on the energy and momentum transfer and are 
essentially due to the different values of the imaginary parts of the two 
potentials, which account for the overall effects of inelastic channels and are not univocally
determined from the elastic phenomenology.
In contrast, the real terms are very similar for different parameterizations and 
give very similar results.
In the rROP calculation shown in Fig.~\ref{f1}, the real part of the EDAI potential has 
been used, but a calculation with
EDAD1 would give in practice the same result. The results in Fig.~\ref{f1} stress the 
importance 
of FSI and, in particular, of the imaginary part of the relativistic optical 
potential, which plays a different role in
the different approaches. In the rROP, the imaginary part is neglected  and the 
total flux is automatically conserved.  
The RGF results presented here contain the contribution of both
terms of the hadron tensor in Eq. (61) of Ref.~\cite{Meucci:2003uy}.
The calculation of the second term, which is entirely due
to the imaginary part of the optical potential, is a hard and time consuming 
numerical task which requires the integration over all the eigenfunctions of the 
continuum spectrum of the optical potential. Numerical uncertainties on this
term are anyhow under control and, from many calculations in different 
kinematics, have been estimated at most within $10\%$.
In the RGF, the imaginary part redistributes the flux in all the final-state 
channels and,  in each channel, the loss of flux
towards other inelastic channels (either multinucleon emission
or non-nucleonic excitations) is compensated for the
inclusive scattering making use of the dispersion relations. 
The larger cross sections in the RGF arise from the translation to the inclusive 
strength of the overall effects of inelastic channels. These contributions are not 
included explicitly in the RGF model, but can be recovered, at least to some extent, in
the RGF by the imaginary part of the phenomenological optical potential 
which reincorporates these processes in the reaction.
\begin{figure}[tb]
\begin{center}
\includegraphics[scale=0.45]{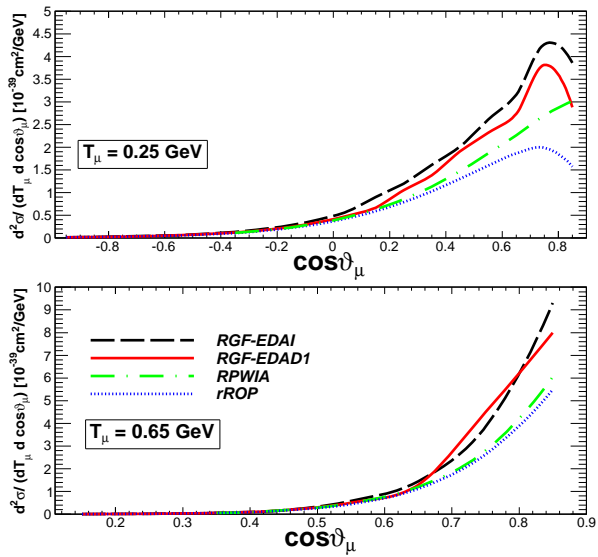} 
\end{center}
	\caption{(color online) Flux-averaged double-differential cross section per target nucleon 
for the CCQE $^{12}$C$\left(\bar\nu_{\mu}, \mu^+\right)$ reaction as a function of 
$\cos\vartheta_{\mu}$ for two bins of $T_{\mu}$ calculated with the RGF-EDAD1 
(solid lines) and the RGF-EDAI (dashed lines). The dotted lines are the rROP results 
calculated with the EDAI potential and the dot-dashed lines are the RPWIA results.
	\label{f2}}
	\end{figure}

In Fig.~\ref{f2} the CCQE flux-averaged double-differential 
$^{12}$C$\left(\bar\nu_{\mu}, \mu^+\right)$ cross section per target 
nucleon is displayed as a function of $\cos\vartheta_{\mu}$ for two bins of $T_{\mu}$. 
Similar considerations can be made for these results: the rROP gives the lower results and
the RGF cross sections with both optical potentials are larger than the RPWIA and the 
rROP ones. 
All the  models, however, tend to produce similar results for
 backward scattering angles, where the cross sections are sensibly reduced.
The differences between the RGF results with the two 
optical potentials are visible but they are somewhat reduced with respect 
to the corresponding calculations for neutrino scattering in Fig.~2 of 
Ref.~\cite{Meucci:2011vd}.
This is essentially due to the combined effects of the differences 
between   the
$\nu_{\mu}$ and the $\bar\nu_{\mu}$ fluxes and of the destructive contribution of 
the  vector-axial interference response.

\begin{figure}[tb]
\begin{center}
\includegraphics[scale=0.45]{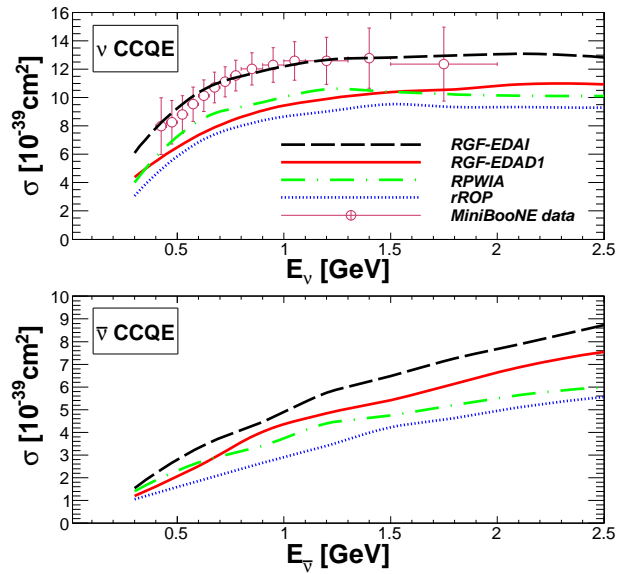} 
\end{center}
	\caption{(color online) Total CCQE cross section per target nucleon as a function of the 
neutrino energy $E_{\nu}$ (upper panel) and of the antineutrino energy $E_{\bar\nu}$ 
(lower panel) calculated with the RGF-EDAD1 (solid lines), the RGF-EDAI (dashed lines), 
the rROP (dotted lines), and the RPWIA (dot-dashed lines). 
The experimental data for neutrino scattering are from MiniBooNE~\cite{miniboone}.
	\label{f3}}
	\end{figure}

In Fig.~\ref{f3} the total QE cross sections per target nucleon for neutrino and 
antineutrino scattering are displayed as a
function of the neutrino (antineutrino) energy $E_{\nu} \left(E_{\bar\nu}\right)$.  
The neutrino results, which 
have been already presented in Ref.~\cite{Meucci:2011vd}, are  compared with the 
experimental data from MiniBooNE~\cite{miniboone}. 
In Refs.~\cite{Amaro:2011qb,Amaro:2011aa} it is shown that  DWIA models where
FSI effects are accounted for by means of real potentials, like rROP and RMF, produce 
similar results which all underestimate the total CCQE MiniBooNE cross section, 
whereas the inclusion of 2p-2h MEC  enhances the results.
Larger cross sections are obtained in the RGF with both optical
potentials. The differences between RGF-EDAI and RGF-EDAD1
are clearly visible, RGF-EDAI being in good
agreement with the shape and magnitude of the experimental
cross section for neutrino scattering. We  observe that EDAI is a 
single-nucleus parameterization, which is 
constructed to better reproduce the elastic proton-$^{12}$C 
phenomenology, and gives electron scattering cross sections in fair 
agreement with the experimental data. 
We also note that the antineutrino cross 
section does not saturate in the energy range up to 2.5 GeV which we have considered. 
Also in this case, the differences between EDAI
and EDAD1 are due to the different imaginary
parts of the two potentials, particularly for the energies
considered in kinematics with the largest $T_{\mu}$. These kinematics
give large contributions to the total cross section and enhance the differences
between the two RGF results.


\section{Summary and conclusions}
\label{conc}

In this paper, we have compared the predictions of different relativistic descriptions 
of FSI for CCQE antineutrino-nucleus scattering in the MiniBooNE kinematics.
In the RPWIA, FSI are simply neglected; in the rROP, they are described retaining only the 
real part of the relativistic energy-dependent optical potential; in the 
RGF, the full complex optical potential, with its real and imaginary parts, is used to
account for FSI. All final-state channels are included
in the RGF, where the flux lost in each channel is
recovered in the other channels  by the imaginary part
of the optical potential making use of the dispersion relations
and the total flux is conserved. The RGF is
able to give a good description of the $(e,e^{\prime})$ data  in the QE region 
and it is also able to describe both the MiniBooNE CCQE and NCE neutrino
data without the need to change the value of the axial mass. 

The enhancement of the RGF cross sections  compared to the cross sections
calculated with other descriptions of FSI is obtained also in the case of antineutrino 
scattering. 
The  larger results of the RGF can be ascribed to the contribution of reaction 
channels which are not included in the other models. For instance, rescattering processes 
of the nucleon in its way out of the nucleus,  non-nucleonic $\Delta$  excitations, which
may arise during nucleon propagation, with or without real
pion production, or also multinucleon processes.
These contributions are not included explicitly in the RGF
model, but can be recovered,
to some extent, by the imaginary part of the
phenomenological optical potential.
We cannot disentangle the role of different reaction processes and explain
in detail the origin of the recovered strength.
It would be anyhow interesting for a comparison to disentangle in the 
phenomenological optical potential the contributions due to non-nucleonic 
inelasticities and extract a \lq\lq purely nucleonic\rq\rq optical potential 
that could be used in the RGF approach.

The RGF predictions are also affected by uncertainties
in the determination of the phenomenological optical potential, which is
not univocally determined from the elastic phenomenology.
In the case of antineutrino scattering the differences between the RGF cross
sections calculated with the EDAI and EDAD1 optical potentials 
are somewhat reduced with respect to the corresponding cross sections calculated
for neutrino scattering. The differences between the RGF results for neutrino 
and antineutrino scattering may be ascribed to the different  $\nu_{\mu}$ and 
$\bar\nu_{\mu}$ fluxes and to the strength of the  vector-axial response.
A better determination of the phenomenological relativistic optical potential, 
which closely fulfills the dispersion relations, would reduce the theoretical
uncertainties on the RGF results and, therefore, deserves further investigation.

The analysis of MiniBooNE CCQE and NCE data
with theoretical  models based on the IA and including only one-nucleon
knockout contributions usually requires a larger value of
$M_A$ to reproduce the magnitude of the experimental cross
sections. The calculations required for the theoretical
analysis must consider the entire range of the
relevant MiniBooNE neutrino energies and additional complications
may arise from the flux-average procedure to
evaluate the cross sections, which implies
a convolution of the double-differential cross section over
the neutrino spectrum. 
Because of uncertainties associated with the flux-average procedure,
the MiniBooNE cross sections can include contributions
from different kinematic regions, where other
reaction mechanisms than one-nucleon knockout are
known to be dominant~\cite{Benhar:2010nx,Benhar:2011wy}. 

Models including other contributions than one-nucleon
knockout, like our RGF, but also the model of Refs.~\cite{Martini:2009uj,Martini:2010ex,Martini:2011wp},
where multinucleon components are explicitly included, are able to describe the MiniBooNE neutrino
data without the need to change the value of the axial mass.
Despite their differences, the two models seem to go in the
same direction. In the RGF, however, the enhancement of
the cross section cannot be attributed only to multinucleon
processes, since we cannot disentangle the role of the
various contributions included in the phenomenological
optical potential.
In order to clarify this point a
careful evaluation of all nuclear effects and of the relevance of multinucleon emission and of some non-nucleonic
contributions would therefore be  highly desirable.

\begin{acknowledgments}

We would like to thank J.A. Caballero and J. Grange for useful comments. 
This work was partially supported by the Italian MIUR 
through the PRIN 2009 research project.

\end{acknowledgments}

%
%


\end{document}